# A new method incorporating deep learning with shape priors for left ventricular segmentation in myocardial perfusion SPECT images

Fubao Zhu[a], Jinyu Zhao[a], Chen Zhao[b], Shaojie Tang[c], Jiaofen Nan[a], Yanting Li[a], Zhongqiang Zhao[d], Jianzhou Shi[d], Zenghong Chen[d], Zhixin Jiang[d]\*, Weihua Zhou[be]\*\*

[a] School of Computer and Communication Engineering, Zhengzhou University of Light Industry, Zhengzhou 450001, China
[b] Department of Applied Computing, Michigan Technological University, Houghton, MI 49931, USA
[c] School of Automation, Xi'an University of Posts and Telecommunications, Xi'an, Shaanxi 710121, China
[d] Department of Cardiology, The First Affiliated Hospital of Nanjing Medical University (Jiangsu Provincial Hospital) Nanjing 210000, China
[e] Center for Biocomputing and Digital Health, Institute of Computing and Cybersystems, and Health Research Institute, Michigan Technological University, Houghton, MI, 49931, USA

\* Corresponding author at: Department of Cardiology, The First Affiliated Hospital of Nanjing Medical University, Jiangsu Province Hospital, 300, Guangzhou Road, Nanjing 210029, China
\*\* Corresponding author at: Department of Applied Computing, Michigan Technological University, Houghton, MI 49931, USA
*E-mail addresses:* zhixin_jiang@njmu.edu.cn (Z. Jiang), whzhou@mtu.edu (W. Zhou)

**Abstract**

**Background:** The assessment of left ventricular (LV) function by myocardial perfusion SPECT (MPS) relies on accurate myocardial segmentation. The purpose of this paper is to develop and validate a new method incorporating deep learning with shape priors to accurately extract the LV myocardium for automatic measurement of LV functional parameters.
**Methods:** A segmentation architecture that integrates a three-dimensional (3D) V-Net with a shape deformation module was developed. Using the shape priors generated by a dynamic programming (DP) algorithm, the model output was then constrained and guided during the model training for quick convergence and improved performance. An MPS dataset, including 31 subjects without or with mild ischemia, 32 subjects with moderate ischemia, and 12 subjects with severe ischemia, were retrospectively analyzed; myocardial contours were manually annotated as the ground truth. A stratified 5-fold cross-validation was used to train and validate our models. Left ventricular end-systolic volume (ESV), end-diastolic volume (EDV), ejection fraction (LVEF), and scar burden were measured from extracted myocardial contours to evaluate the clinical performance.
**Results:** Results of our proposed method agree well with those from the ground truth. Our proposed model achieved a Dice similarity coefficient (DSC) of 0.9573±0.0244, 0.9821±0.0137, and 0.9903±0.0041, a Hausdorff distances (HD) of 6.7529±2.7334 mm, 7.2507±3.1952 mm, and 7.6121±3.0134 mm in extracting the endocardium, myocardium, and epicardium, respectively. The correlation coefficients between LVEF, ESV, EDV, stress scar burden, and rest scar burden measured from the optimal model in the stratified 5-fold cross-validation and those from the ground truth were 0.92, 0.958, 0.952, 0.972, and 0.958, respectively.
**Conclusion:** Our proposed method achieved a high accuracy in extracting LV myocardial contours and assessing LV function. Future studies will further improve our method and investigate the values of LV parameters measured from our method in a clinical environment.

# 1. Introduction

Coronary artery disease (CAD) is currently one of the diseases with the highest morbidity and mortality in the world [1]. Gated myocardial perfusion SPECT (MPS) is widely used for non-destructive diagnosis of CAD due to its excellent efficacy/cost ratio in assessing left ventricular function [2]. For quantitative analysis of the left ventricle (LV) in MPS, the endocardium, myocardium, and epicardium must be accurately delineated on perfusion images, followed by measurement of LV functional parameters [3]. Manual segmentation is time-consuming and lacks reproducibility [6]. Therefore, it is quite necessary to develop a precise, reproducible, and fully automated segmentation algorithm to improve the accuracy of quantitative analysis.

Currently, commercial software extracts the endocardial and epicardial surfaces by identifying the maximum myocardial counts and then applies a Gaussian fit with empirical standard deviation or threshold method to estimate the myocardial profile [7]. However, this method produces errors in assessing myocardial functions. In particular, left ventricular ejection fraction (LVEF) is often overestimated in patients with small hearts, and the error is more pronounced in female than in male [9].

Traditional image processing techniques have demonstrated good performance in cardiac image segmentation, such as atlas-based methods and model-based methods [12-15]. In recent years, deep learning models, which automatically learn high-level features of the potential distribution of data, have already outperformed traditional image segmentation algorithms in both accuracy and time efficiency [16]. Wang et al. [17] proposed the model by a multi-class three-dimensional (3D) V-Net to automatically extract the endocardium and epicardium in gated MPS, which exhibits excellent segmentation performance. The average Dice similarity coefficients (DSC) values of the model in the endocardium, myocardium, and epicardium of normal patients were 0.907, 0.926 and 0.965, respectively. Hausdorff distance (HD) for the endocardium was 8.402 mm and for the epicardium was 8.631 mm. Wen et al. [18] generated a preliminary dataset based on the dynamic programming (DP) tool and then manually adjusted the data with significant errors before using a U-Net network to extract the myocardial contours. The DSCs of this model on gated MPS images for endocardium, myocardium, and epicardium segmentation were 0.9222, 0.9580, and 0.9748, respectively. The HDs were 7.4767 mm, 7.7911 mm, and 8.0003 mm, respectively. Nevertheless, these deep learning methods still need to be improved for accurate LV segmentation in MPS. The object shapes extracted from traditional image segmentation algorithms have shown great success as prior knowledge in refining the deep learning models for medical image segmentation [19]. Combining the prior knowledge reduces the potential output space of model partitioning and speeds up the convergence during the model training [22]. However, prior knowledge is generally used as the model's input and is not easily obtained.

This study proposed a new deep learning-based method, called dynamic-programming shape-transformation V-net (DP-ST-V-Net). The innovations and our contributions are listed as follows:

(1) The myocardial contours generated by a DP algorithm were incorporated into the network as the prior knowledge, which significantly improved the accuracy of LV segmentation results.

(2) A novel segmentation architecture was proposed, which integrated a spatial transform network (STN) into a 3D V-Net to constrain and guide the model for segmentation. By deforming the shape prior, reliable segmentation results were produced.

# 2. Materials and methods

The proposed segmentation process is shown in Fig. 1. Firstly, the MPS image and the shape prior obtained by a DP algorithm are used as the dual-channel input of the V-Net to generate the coarse segmentation results, as shown in step 1. Then the output of the V-Net is used as the input of a STN, and the

STN network is trained to improve the segmentation results according to the shape prior generated by the DP algorithm, as shown in step 2. Finally, the V-Net and STN are combined to train the entire DP-ST-V-Net network, as shown in step 3.

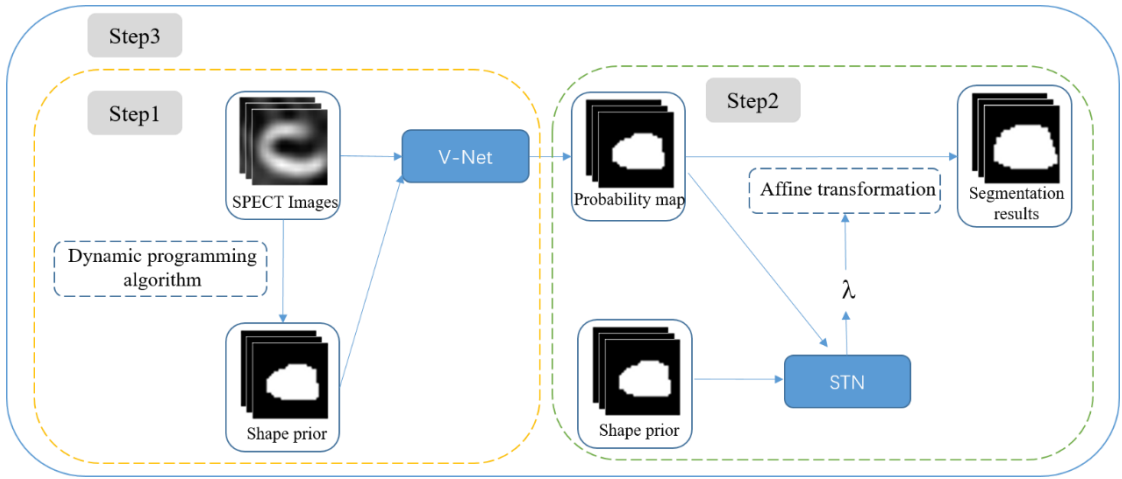

**Fig. 1.** Proposed workflow for DP-ST-V-Net. The network consists of a V-Net and a STN. V-Net is used to extract the LV contours, and STN improves the V-Net output results. Shape prior generated by DP is used to restrict the search space and optimize the results. λ is the spatial transformation parameter generated by STN, and the probability map is non-rigidly transformed according to λ.

## 2.1 Image acquisition and pre-processing

Seventy-five patients (47 males and 28 females, age 38 to 83 years, mean age 62.18±11.67 years old) were retrospective enrolled. They underwent MPS from March 2014 to August 2017 in the Jiangsu Provincial Hospital, China. The study was approved by the hospital's Medical Ethics Committee, and all patients signed an informed consent form.

Each patient underwent 8 frames of ECG-gated rest and stress SPECT imaging after the injection of 20-30 mCi Tc-99m sestamibi. Thus, a total of 1200 3D MPS image volumes (75 patients ×2 [rest or stress] ×8) with a voxel size of 6.4 mm × 6.4 mm × 6.4 mm were available for this study. Patients were classified into three categories by clinical cardiologists based on medical conditions and MPS images at the enrollment: 31 subjects without ischemia or with mild ischemia, 32 subjects with moderate ischemia, and 12 subjects with severe ischemia.

During data preprocessing, each 3D MPS image volume was first cut longitudinally to generate 32 2D long-axis slices, as shown in Fig. 2 [24], and then each slice was automatically cropped into 32 × 32 pixels. In addition, data augmentation, including random scaling, flipping, and rotation, was performed on the 3D volume to increase the sample size and reduce overfitting during model training. The endocardium, myocardial, and epicardium were manually annotated by experienced operators and confirmed by clinical cardiologists, and then used as the ground truth.

The dataset including de-identified MPS images and manual annotations is publicly available at https://github.com/MIILab-MTU/SPECTMPISeg. A 5-fold cross validation method was used to train our model and evaluate the segmentation results. For each fold, four-fifths of the data was used for training and the other fifth is used for testing. To ensure the consistency of the training and test set, stratified random sampling was performed so that subjects with different ischemia conditions were randomly selected according to their proportions. For fair comparison, both our proposed method and benchmark models were trained and validated using the same data.

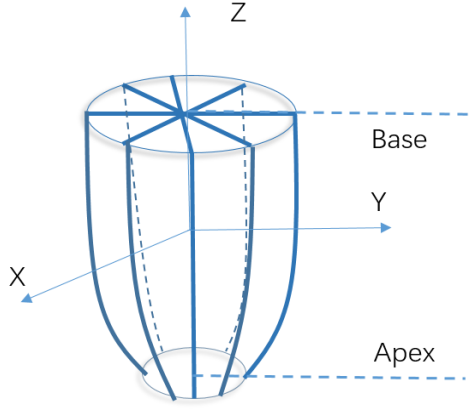

**Fig. 2.** MPS image volume is converted into long-axis slices by cutting it along the longitudinal direction of the left ventricle.

## 2.2 Shape prior generation

The automatic generation of shape prior information is a complex problem. Even if acquired, inaccurate shape priors may mislead the model and make it collapse during the training process. Compared with machine learning methods, the most obvious advantages of the DP algorithm are the simplicity and efficiency. In the DP procedure [24], the MPS images are first transformed from the original Cartesian coordinates to polar coordinates. Then the endocardial and myocardial contours are initially determined using the DP algorithm. After correcting the LV valve plane, the endocardial and epicardial contours are also determined. Relevant studies have proved that this method has achieved good results in the LV myocardium segmentation using gated MPS images [24]. Therefore, the myocardial contours generated by the DP algorithm are suitable for generating the shape priors in this study.

## 2.3 Proposed DP-ST-V-Net model for LV myocardial segmentation

The DP-ST-V-Net consists of a 3D V-Net for extracting myocardial contours and a shape deformation module to improve the output of the V-Net, as shown in Fig. 3.

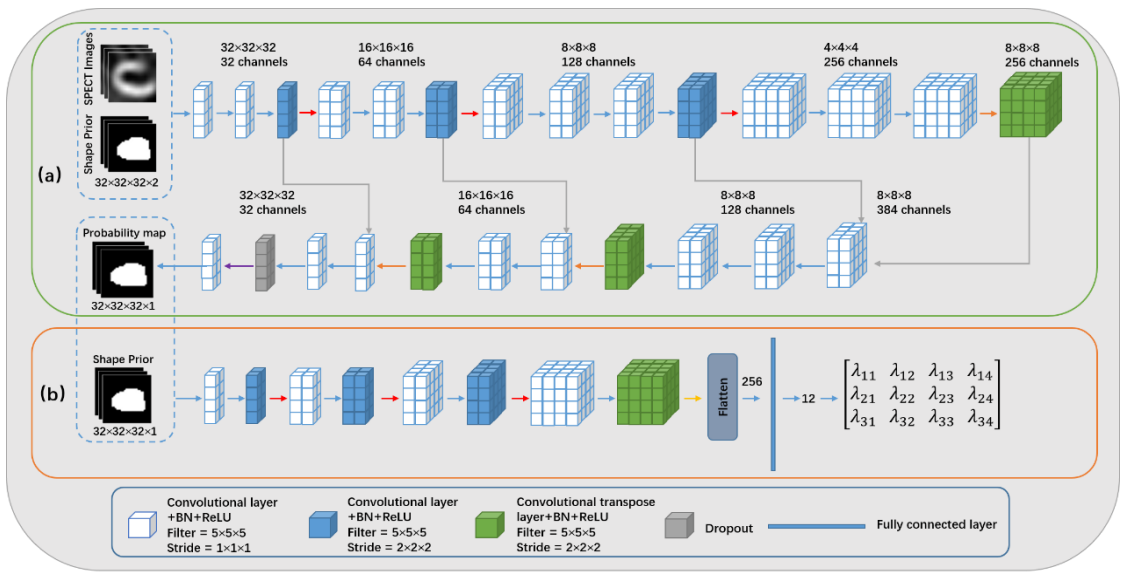

**Fig. 3.** The proposed DP-ST-V-Net model. (a), end-to-end segmentation of the LV myocardial contour. (b) Optimization of the V-Net output results.

### 2.3.1 The proposed V-Net module

The V-Net in this work is an end-to-end volume-based 3D image segmentation method, as shown in Fig. 3(a). The original SPECT image and the binary shape prior mask are first cropped to a cubic with the size of 32 × 32 × 32 pixels to prevent them from being affected by the background area and then simultaneously input into the V-Net. The model's output is a probability map of the LV myocardial contour with the same size as the original image.

In detail, the V-Net consists of an encoder that extracts low-level and high-level features and a decoder that restores features. Both the encoder and the decoder contain multiple stages. As the image goes through the network, its resolution first decreases and then increases. The encoder is divided into multiple stages, and each stage contains two to three convolutional layers. The convolutional layer in the white block contains a kernel with a size of 5 × 5 × 5 and a stride of 1, resulting in the same resolution of the output feature maps but increasing the number of channels. Then, a convolutional layer in the blue block contains a kernel with a size of 2 × 2 × 2, and a stride of 2 is employed to reduce the resolution by half. After performing the convolution, the resolution of the feature maps is doubled using the deconvolution layers. Finally, a dropout layer [25] is added before the last convolutional layer, and the probability value is set to 0.8 to avoid over-fitting.

### 2.3.2 Shape deformation module

While the segmentation results generated by the DP algorithm have already achieved high performance, several predicted contours are not completely falling within the ideal position. Especially for some small heart patients, their valve planes cannot be generated accurately using DP algorithms [26]. However, inputting shape prior information generated by DP into V-Net reduces segmentation accuracy and affects the model performance. To address this problem, the shape deformation module is used to restrict the output of V-Net based on shape prior information to improve the accuracy of segmentation results.

The shape deformation module is implemented by STN [27], as shown in Fig. 3(b). It utilizes the automatically learned spatial transformation parameters to spatially align the shape prior into the ground truth, reducing the impact of the image on the segmentation task due to the spatial transformation. The objective function includes the difference between the transformed images, and the corresponding ground truth is utilized to optimize the network weights. The input of the STN is the V-Net segmentation result and the generated shape prior. They form a dual-channel feature map with the size of 32 × 32 × 32 × 2. The output of the STN model is 12 parameters, representing the non-rigid affine transformation, as defined in λ. The whole shape deformation module contains 7 convolutional layers for extracting image features. The Flatten layer converts the extracted feature maps into a one-dimensional feature vector, which is then generated by the fully connected layer into a corresponding transformation matrix.

In the DP-ST-V-Net model, the original SPECT image is denoted as $S$, and the shape prior generated by DP is denoted as $D$. The manually delineated myocardial contour is denoted as $Y$, the V-Net output is denoted as $Y'$, the result after performing the affine transformation is denoted as $Y''$, and their sizes are of $L \times W \times H$, where $L$, $W$ and $H$ represent the length, width, and height of the input image, respectively. The output result $Y'$ of V-Net and the shape prior $D$ form a two-channel image as STN input. The output of the STN model is the parameters of the affine transformation matrix λ.

The STN is divided into three parts: localization net, grid generator, and sampler.

The localization net $f_{loc}$ is used to generate the transformation parameters $\lambda_{x_i, y_i, z_i}$, $x, y, z$, representing the coordinates of each pixel in the image. The images $Y'$ and $D$ are the input of the localization network. After performing the convolution and flattening the feature maps, the spatial transformation parameters are generated as follows:

$$\lambda_{x_i,y_i,z_i} = f_{loc}(Y',D) \qquad (1)$$

The grid generator constructs a sampling grid based on the transform parameters predicted by the localization net. The input image is passed through the grid generator to obtain the affine transform function $T_{\lambda_{x_i,y_i,z_i}}$. Assuming that each pixel coordinate of the feature image is $(x_i^s, y_i^s, z_i^s)$, each pixel coordinate of the target image $Y''$ is $(x_i^t, y_i^t, z_i^t)$, and $t$ denotes the output target, the correspondence between $(x_i^s, y_i^s, z_i^s)$ and $(x_i^t, y_i^t, z_i^t)$ is obtained as

$$\begin{pmatrix} x_i^s \\ y_i^s \\ z_i^s \end{pmatrix} = T_{\lambda_{x_i,y_i,z_i}}(G) = A_{\lambda_{x_i,y_i,z_i}} \begin{pmatrix} x_i^t \\ y_i^t \\ z_i^t \\ 1 \end{pmatrix} = \begin{bmatrix} \lambda_{11} & \lambda_{12} & \lambda_{13} & \lambda_{14} \\ \lambda_{21} & \lambda_{22} & \lambda_{23} & \lambda_{24} \\ \lambda_{31} & \lambda_{32} & \lambda_{33} & \lambda_{34} \end{bmatrix} \begin{pmatrix} x_i^t \\ y_i^t \\ z_i^t \\ 1 \end{pmatrix}, \quad i \in [1,\cdots,L\times W\times H], \qquad (2)$$

where $A_{\lambda_{x_i,y_i,z_i}}$ is the affine relationship, and the output of $G$ is the target pixel location.

Based on the above processing results, the sampling transformation of each coordinate point of the output feature map is completed. Cubic spline interpolation is used to calculate the sampling coordinates to prevent overflow of the coordinate positions, so that the coordinates of the voxels fall within the integer grid, as shown in Eq. (3),

$$Y_i^{''n} = \sum_l^L \sum_w^W \sum_h^H Y_{lwh}^{'n} \max(0, 1-|x_i^s - l|) \max(0, 1-|y_i^s - w|) \max(0, 1-|z_i^s - h|),$$

$$\forall i \in [1,\cdots,L\times W\times H], \qquad (3)$$

where $n$ represents the number of channels of the feature map, and $Y_{lwh}^{'n}$ is the coordinate of the position $(L, W, H)$ of the input feature map $Y'$ at the channel $n$. The STN, composed of the above three parts, which can be inserted into the neural network independently and trained jointly in the network to modify the parameters to complete the affine transformation of the feature information.

**2.4 Loss function and optimization**

**2.4.1 Loss function**

DP-ST-V-Net contains a V-Net network segmenting the LV myocardial contour and a shape deformation module. In order to optimize the weight of each network and improve the model's performance, we employ three loss functions in Fig. 4.

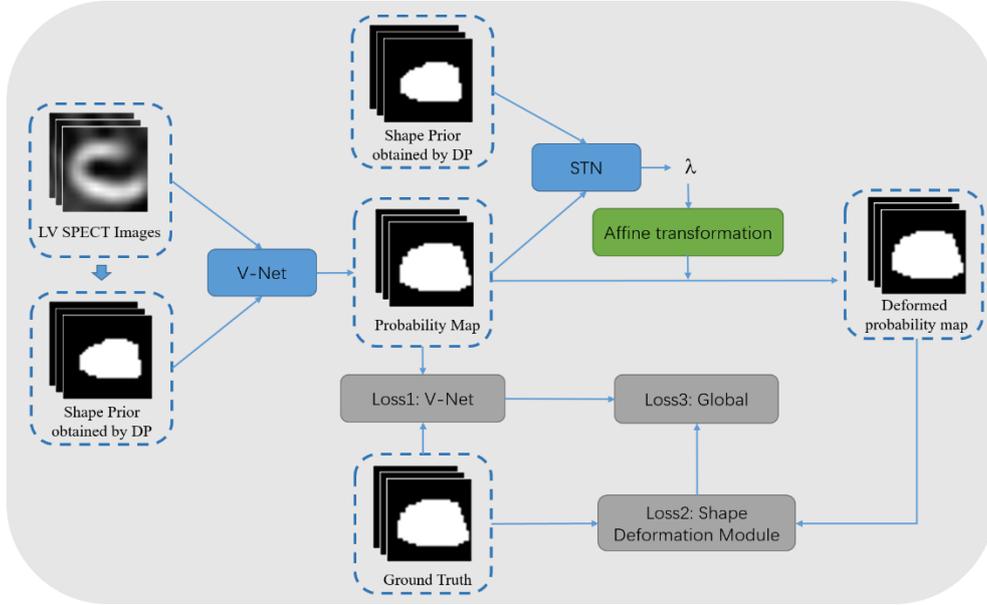

**Fig. 4.** Loss function for DP-ST-V-Net.

1. **Loss function for V-Net**

To train the V-Net network, a sigmoid cross entropy loss function is used to minimize the difference between the output $Y'$ and the ground truth $Y$ of the V-Net network, as shown in Fig. 4 for the loss 1. Suppose the activation function of ground truth $Y$ is $V$, then the loss function of V-Net is designed as follows

$$V = sigmoid(Y_{lwh}),$$

$$L_{V-Net} = \sum_{l}^{L}\sum_{w}^{W}\sum_{h}^{H}\left(-Y'_{lwh}\log V - (1-Y'_{lwh})\times\log(1-V)\right) + \|W_1\|_2, \quad (4)$$

where $W_1$ represents all the weights in the V-Net network, and $\|W_1\|_2$ represents the $L_2$ norm of $W_1$ for regularization.

2. **Loss function for shape deformation module**

The gold standard is unknown for affine transformation parameters output by STN networks. Therefore, in this study, the STN output parameters λ are used to perform the affine transform using the output $Y'$ of V-Net. Then the weight of the shape deformation module is finely optimized by the difference between the transformed probability image $Y''$ and the ground truth $Y$, as shown in Fig. 4 for the loss 2. Another sigmoid cross entropy loss function minimizes the difference between $Y''$ and $Y$ as shown in Eq. (5),

$$V = sigmoid(Y_{lwh}),$$

$$L_{Deformation} = \sum_{l}^{L}\sum_{w}^{W}\sum_{h}^{H}\left(-Y''_{lwh}\log V - (1-Y''_{lwh})\times\log(1-V)\right) + \|W_2\|_2, \quad (5)$$

where $W_2$ represents the ownership weight of the shape deformation module.

### 3. Loss function for overall objective

Finally, to train the global network, the V-Net network and the shape deformation module are jointly trained, as the loss3 in Fig. 4, with a loss function defined in Eq. (6),

$$V = sigmoid(Y_{lwh}),$$

$$L_{Global} = AL_{V-Net} + BL_{Deformation} + C\|W\|$$

$$= \sum_{l}^{L}\sum_{w}^{W}\sum_{h}^{H}\left(-Y_{lwh}^{"}\log V - (1-Y_{lwh}^{"}) \times \log(1-V) - Y_{lwh}^{"}\log V - (1-Y_{lwh}^{"}) \times \log(1-V)\right) + \|W_1\|_2 + \|W_2\|_2 \quad (6)$$

where $\|W\|$ represents all the weights of the DP-ST-V-Net network. $A$, $B$, and $C$ are the hyperparameter balanced V-Net, shape transformation module, and $L_2$ regularization, respectively.

The V-Net and shape deformation module contain 7.4 and 1.21 million weights, respectively. An $L_2$ regularization term is added to each objective function to prevent model overfitting and accelerate convergence.

### 2.4.2 Optimization strategy

To improve the model's performance, a multi-step training strategy is designed, as shown in Algorithm 1. Eq. (4) is used as the loss function in the first stage. During this period, only the V-Net network is optimized. The shape deformation module is randomly initialized and has a fixed weight value. Eq. (5) is used as the loss function in the second stage. V-Net weights are fixed, and only the shape deformation module is trained. In the third stage, DP-ST-V-Net is trained jointly using Eq. (6) as the loss function. During the training period, Adam Optimizer is used to optimize the network structure, with a total training epoch of 3000 and a batch size of 8.

**Algorithm 1**
DP-ST-V-Net multi-step training strategy

Input: $S$: Cropped SPECT raw images; $Y$: Manual depiction of the LV myocardial contour;
  $e$: Number of epochs during training;
Output: DP-ST-V-Net for training;
Step1: Epoch is set to $0.1e$-$0.2e$, V-Net is trained using Eq. (4) as the loss function;
Step2: Epoch is set to $0.2e$-$0.3e$, shape deformation module is trained using Eq. (5) as the loss function;
Step3: Epoch is set to $0.3e$-$e$, DP-ST-V-Net is trained using Equation (6) as the loss function.

### 2.5 Evaluation metrics

The accuracy of the method is first quantitatively analyzed for measuring the performance of the model segmentation by the well-known evaluation metrics [28], including DSC and HD.

DSC is used to measure the overlap between the model segmentation foreground and the ground truth. If the region predicted by the model is denoted by $Y'$ and the ground truth region is denoted by $Y$, the DSC is defined in Eq. (7).

$$DSC = \frac{2|o(Y') \cap Y|}{|o(Y')| + |Y|} \tag{7}$$

where $o(Y')$ represents the binarization of the model prediction. The higher the ratio of the overlapping, the closer DSC is to 1. A higher DSC indicates a better model performance.

HD is the maximum distance used to measure the set of the shortest distances between the model segmentation foreground and the ground truth [30]. A smaller HD value indicates a more remarkable similarity between the predicted result and the ground truth. HD is defined in Eq. (8) and Eq. (9):

$$HD(Y',Y) = \max(h(Y',Y), h(Y,Y')) \tag{8}$$

where function $h(A,B)$ represents:

$$h(A,B) = \min(\max(\|a-b\|)), a \in A, b \in B \tag{9}$$

To further measure the accuracy of DP-ST-V-Net segmentation, clinical parameters (including myocardial volume, ESV, EDV, LVEF, and scar burden) were calculated using the segmentation results of DP-ST-V-Net, and compared with the ground truth results as well as the results of the commercial software (Emory Cardiac Toolbox 4.0; Atlanta, GA).

The area of the myocardial endocardial surface is generally considered to be the LV cavity. The minimum and maximum endocardial volumes among all phases (0, 1/8... 7/8) represent ESV and EDV, respectively. After obtaining the endocardial surfaces, ESV and EDV are calculated using solid geometry. In particular, the calculated sampling points of ESV and EDV using our method are consistent with the commercial software. LVEF is defined in Eq. (10) [31]:

$$EF = \frac{EDV - ESV}{EDV} \tag{10}$$

For quantitative and qualitative assessment of scar burden, the myocardial perfusion distribution is constructed from the myocardium trajectory. The number of abnormal pixels (below the normal threshold for the location of the myocardium is then determined and proportional to the number of pixels in the entire myocardial region so that scar burden can be expressed as a percentage value.

Furthermore, the correlation degree of clinical parameters between DP-ST-V-Net and the ground truth is analyzed by Pearson correlation analysis, and the correlation coefficient (r) and P-value were calculated. A correlation coefficient r value closer to 1 represents a higher accuracy of DP-ST-V-Net. When the P value<0.05, it is considered as being statistically significant. The relative error of the DP-ST-V-Net measurement of myocardial volume shows the magnitude of its deviation from the ground truth result, and its corresponding formula is defined as Eq. (11):

$$\delta = \Delta / L \times 100\% \tag{11}$$

where $\delta$ represents the actual relative error, $\Delta$ represents the absolute error, and $L$ represents the ground truth.

Corresponding Bland-Altman plots are drawn to show the absolute error between DP-ST-V-Net and the ground truth and the relationship between the error and myocardial volume size. At the same time, other clinical indicators are calculated in the same method as above to verify the accuracy of DP-ST-V-Net segmentation results.

## 3. Experimental results

**Table 1**
Comparison of different segmentation methods on MPS images.

| Method | Authors | Patient type | DSC Endo | DSC Myo | DSC Epi | HD (mm) Endo | HD (mm) Myo | HD (mm) Epi |
|---|---|---|---|---|---|---|---|---|
| 3D V-Net | Wang et al. [17] | Normal | 0.907 | 0.926 | 0.965 | 8.402 | N/A | 8.631 |
|  |  | Abnormal | 0.910 | 0.927 | 0.965 | 8.384 | N/A | 9.310 |
| 3D U-Net | Wen et al. [18] | All | 0.9222 | 0.9580 | 0.9748 | 7.4767 | 7.7911 | 8.0003 |
| DP-ST-V-Net |  | All | **0.9573** | **0.9821** | **0.9903** | **6.7529** | **7.2507** | **7.6121** |

*Endo* Endocardium, *Myo* Myocardium, *Epi* Epicardium

Table 1 shows the results of the different segmentation methods in the MPS images. Two baseline models were implemented to demonstrate the validity of DP-ST-V-Net by removing the shape deformation module and shape prior information, respectively. The experimental results are shown in Table 2. Thus, a single-channel (MPS image) input V-Net model (V-Net) was obtained, which removes the shape deformation module and shape prior. The other is a V-Net model (MC-V-Net) with dual-channel (MPS image and shape prior) inputs, which was generated by removing the shape deformation module. Eq. (4) was used as the loss function for V-Net and MC-V-Net during the training process. During the experiments, DP-ST-V-Net, V-Net, and MC-V-Net used the same V-Net structure. Table 2 gives the DSC, HD, training time and test time results of these four segmentation methods for endocardium, myocardium and epicardium segmentation in 75 patients. The time taken to process a 3D MPS volume at a phase (prediction size of 32 × 32 × 32) is used as the test time.

**Table 2**
Mean DSC, HD, training time (for the training dataset) and test time (for each 3D volume) from 75 patients using DP, V-Net, MC-V-Net, and DP-ST-V-Net segmentation methods.

| Metrics | Method | Endocardium | Myocardium | Epicardium |
|---|---|---|---|---|
| DSC | DP | 0.8043 ± 0.1265 | 0.9151 ± 0.0741 | 0.9345 ± 0.0375 |
|  | V-Net | 0.9258 ± 0.0273 | 0.9420 ± 0.0171 | 0.9633 ± 0.0119 |
|  | MC-V-Net | 0.9347 ± 0.0264 | 0.9716 ± 0.0148 | 0.9845 ± 0.0109 |
|  | DP-ST-V-Net | **0.9573 ± 0.0244** | **0.9821 ± 0.0137** | **0.9903 ± 0.0041** |
| HD (mm) | DP | 8.4138 ± 4.9337 | 9.0512 ± 5.5978 | 9.3201 ± 5.8443 |
|  | V-Net | 7.5732 ± 3.6824 | 7.9514 ± 3.5663 | 8.2439 ± 4.8013 |
|  | MC-V-Net | 6.8493 ± 2.8815 | 7.4190 ± 3.2462 | 7.8596 ± 3.1405 |
|  | DP-ST-V-Net | **6.7529 ± 2.7334** | **7.2507 ± 3.1952** | **7.6121 ± 3.0134** |
| Training time (hours) | DP | N/A | N/A | N/A |
|  | V-Net | **32.1869 ± 0.1169** | **32.3561 ± 0.1173** | **32.5595 ± 0.1193** |
|  | MC-V-Net | 48.0532 ± 0.1473 | 48.1467 ± 0.1485 | 48.1604 ± 0.1487 |
|  | DP-ST-V-Net | 57.9210± 0.1822 | 58.0313 ± 0.1832 | 58.0659 ± 0.1848 |
| Test Time (seconds) | DP | 3. 6572 ± 0.3349 | 5.7138 ± 0.4176 | 7.4812 ± 0.4912 |
|  | V-Net | **0.2304 ± 0.0002** | **0.2318 ± 0.0002** | **0.2320 ± 0.0003** |
|  | MC-V-Net | 3.9724 ± 0.3352 | 6.0029 ± 0.4178 | 7.6635 ± 0.4915 |
|  | DP-ST-V-Net | 4.5899 ± 0.3352 | 6.6463 ± 0.4179 | 8.4202± 0.4916 |

As shown in

Table **3**, for DP-ST-V-Net, all the mean DSC values of the three types of myocardial ischemia severity are greater than 0.95, and the Hausdorff distances are less than 8 mm, which is less than 1.5 voxels since the voxel size is 6.4*6.4*6.4 mm$^3$. These results quantitatively show that the proposed DP-ST-V-Net achieve a better performance compared to the baseline models.

**Table 3**

Mean DSC and HD values of DP-ST-V-Net in endocardium, myocardium, and epicardium with different severity of myocardial ischemia.

| Metrics | Severity | Endocardium | Myocardium | Epicardium |
|---|---|---|---|---|
| DSC | Normal or Mild | 0.9582 ± 0.0216 | 0.9846 ± 0.0114 | 0.9915 ± 0.0034 |
| | Moderate | 0.9571 ± 0.0223 | 0.9854 ± 0.0138 | 0.9894 ± 0.0045 |
| | Severe | 0.9563 ± 0.0251 | 0.9821 ± 0.0169 | 0.9872 ± 0.0063 |
| HD (mm) | Normal or Mild | 6.6855 ± 2.4438 | 7.2412 ± 3.0174 | 7.5360 ± 2.7543 |
| | Moderate | 6.7221 ± 2.6492 | 7.2392 ± 3.2146 | 7.6473 ± 2.9492 |
| | Severe | 6.7859 ± 2.8432 | 7.2721 ± 3.2264 | 7.6878 ± 3.1277 |

The segmentation results of DP-ST-V-Net, V-Net, MC-V-Net and DP are shown in Fig. 5 (One patient with mild or normal, one patient with moderate and one patient with severe myocardial ischemia were selected). Fig. 6 shows the segmentation results with moderate myocardial ischemia in 4 different gated phases. The slices at the same location are selected in each gated phase.

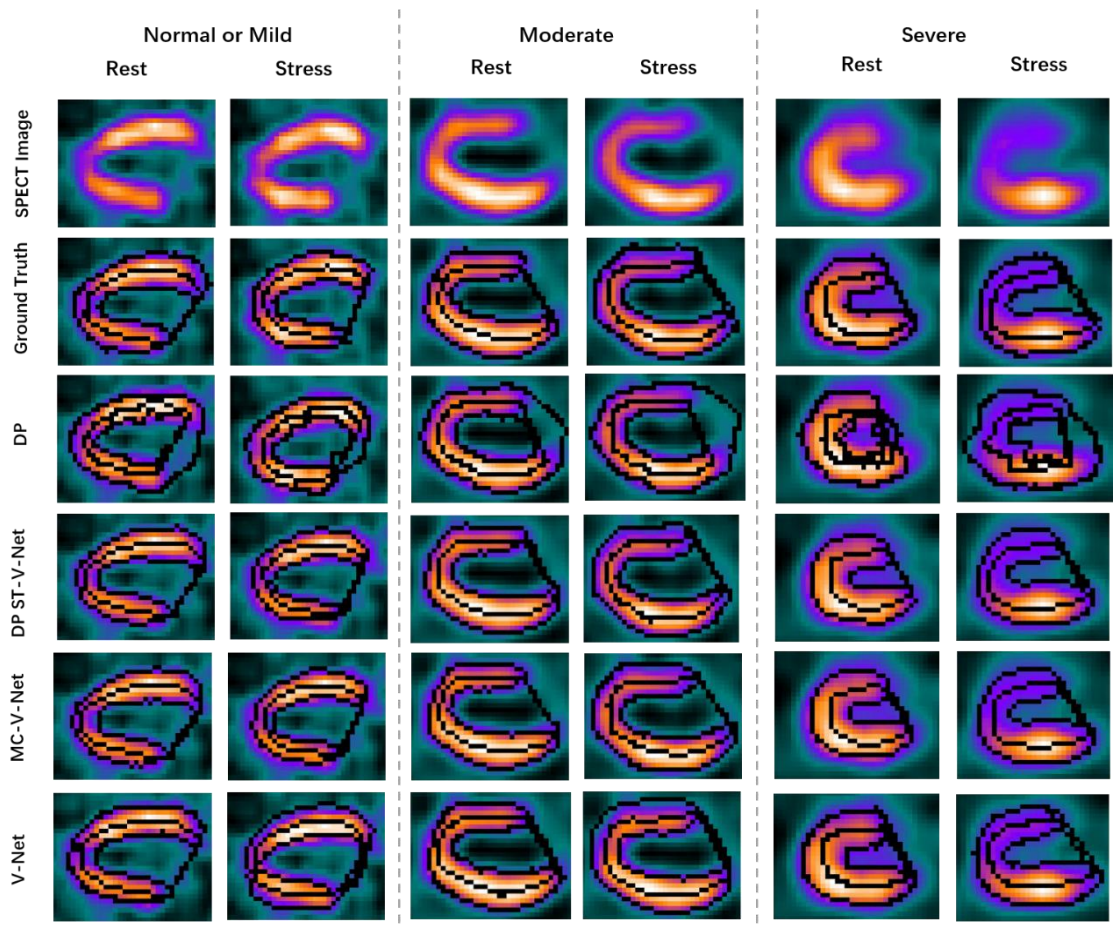

**Fig. 5.** Segmentation result images. From left to right are one normal or mild, one moderate, and one severe myocardial ischemia patient. The first row shows the original MPS image; the second row shows the ground truth; the third row shows the myocardial contours obtained by the DP; and in rows 4-6, the myocardial contours were obtained by DP, DP-ST-V-Net, MC-V-Net and V-Net, respectively.

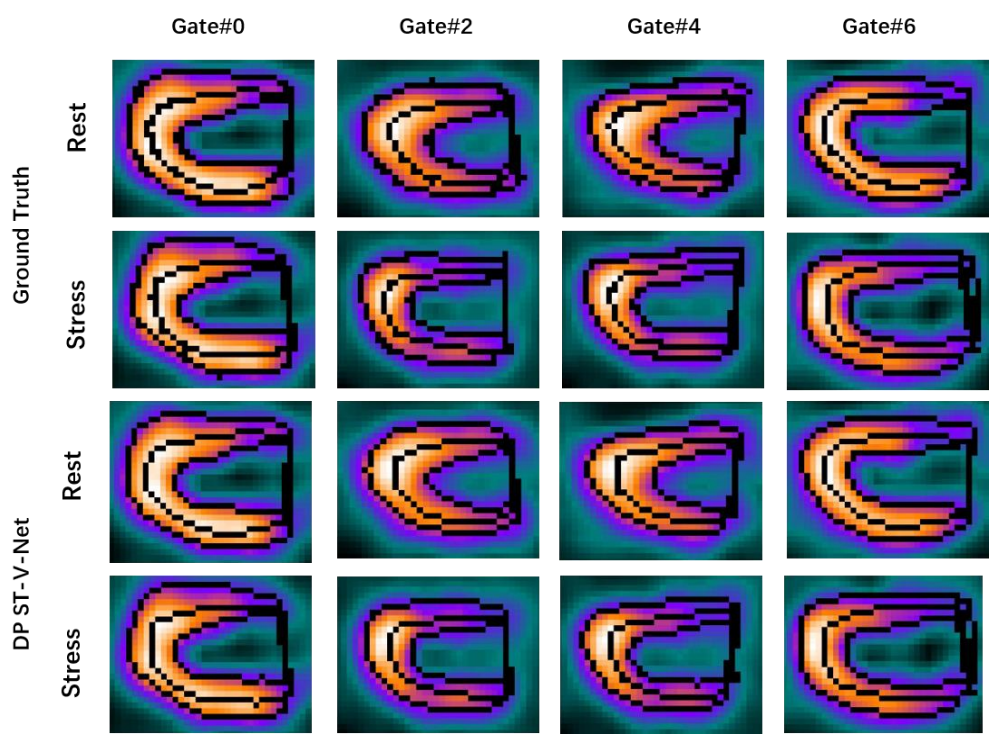

**Fig. 6.** Images of segmentation results for a patient with moderate disease. The first two rows show the manually outlined myocardial contours at four gated phases during loading and resting. The last two rows

are the segmentation results of DP-ST-V-Net at four gated phases during the loading and resting states, respectively.

Fig. 7 shows that the myocardial volume of a patient calculated by the four segmentation methods during one cardiac cycle is shown. The myocardial volume measured by DP-ST-V-Net is 169.58cc, which is underestimated by 0.685% compared to the 170.75cc measured by the ground truth. DP-ST-V-Net is a more accurate reflection of myocardial volume than the baseline model. Notably, data from other patients are compared in the same method, and experimental results are similar. Therefore, this method can accurately quantify changes in myocardial volume within a cardiac cycle.

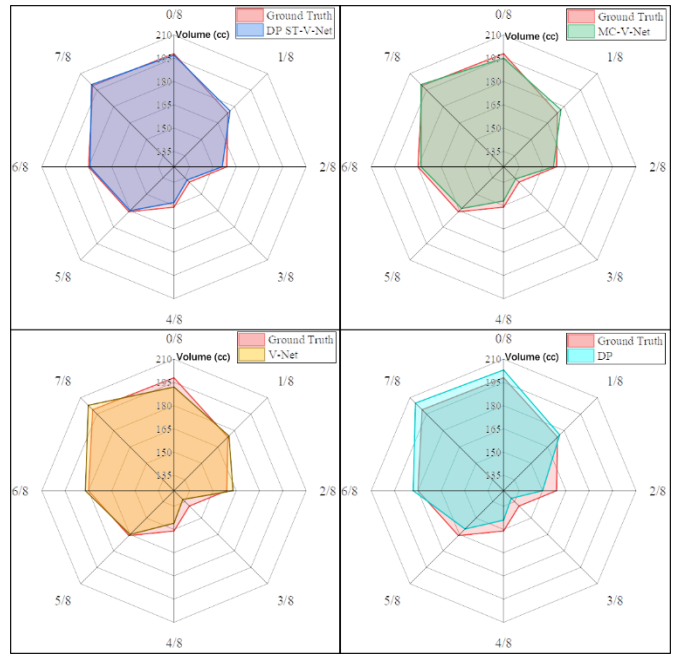

**Fig. 7.** Correlated changes in LV myocardial volume measured at different phases using four segmentation methods in one patient are compared with ground truth myocardial volume.

Correlation analysis of LV volume measured by DP-ST-V-Net and the ground truth during different phases, the calculated r values are shown in Fig. 8. The r value for each phase is greater than 0.89 and the P-value is less than 0.001, indicating a significant linear correlation between the DP-ST-V-Net and the ground truth. Correlation analysis of the segmentation results of DP-ST-V-Net with ground truth in Gate#0 and Gate#3 are shown in Fig. 9, and the relative error of the corresponding LV volume is plotted as a Bland-Altman plot. The mean correlation coefficient between the calculated LV myocardial volume and the ground truth in all patients for each phase is $0.9434 \pm 0.0339$, with a mean relative error of $-0.86 \pm 2.88\%$. The correlation coefficient between the volume size and the volume error for all phase measurements is 0.187 (P=0.262).

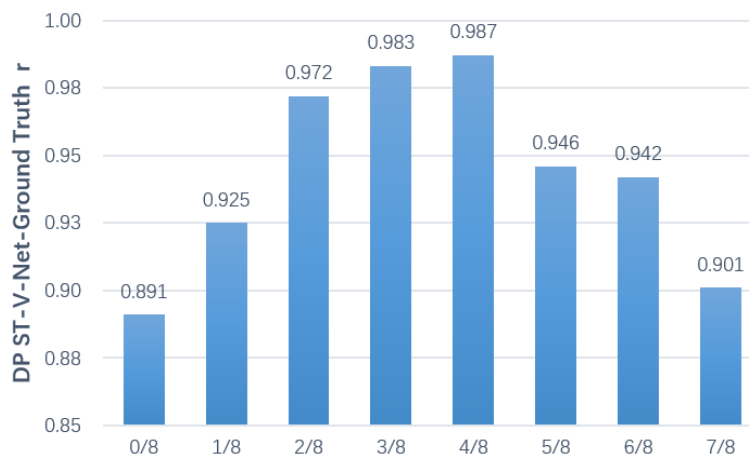

**Fig. 8.** R values are calculated by DP-ST-V-Net and ground truth for measuring the LV myocardial volume in each phase.

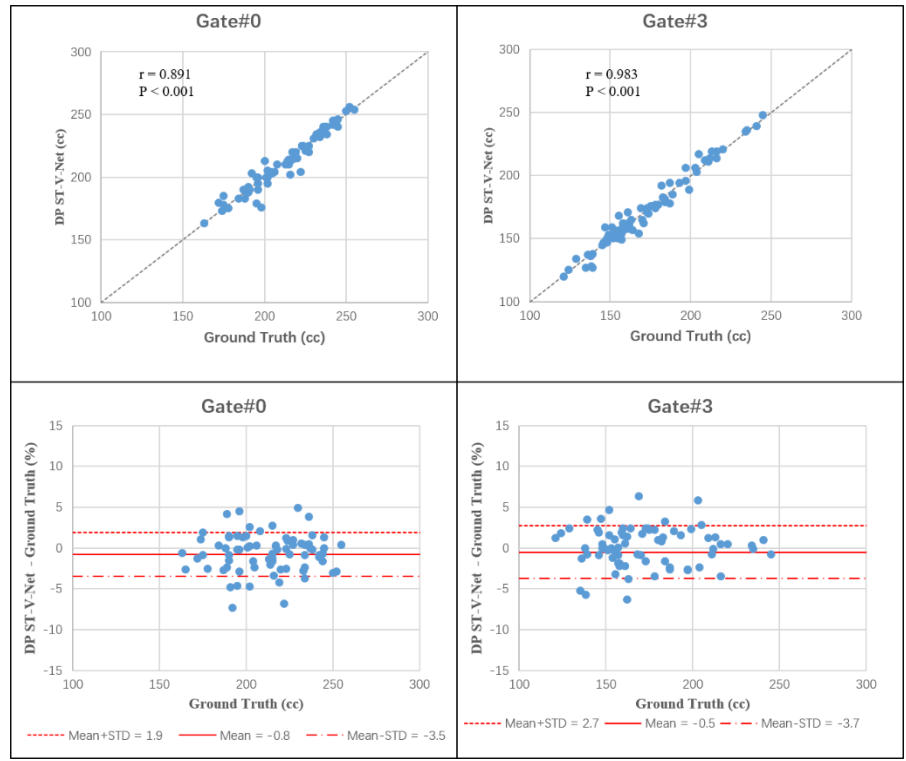

**Fig. 9.** Correlation analysis and relative error between DP-ST-V-Net and ground truth measurement of LV myocardial volume at Gate#0 and Gate#3.

To verify the feasibility of the DP-ST-V-Net method in evaluating clinical parameters, the segmentation results were used to calculate LVEF, ESV, and EDV, and the correlation analysis was performed and compared with ground truth and commercial software results, as listed in Table 4. Fig. 10 shows the results of the correlation analysis of LVEF between the four segmentation methods and ground truth and professional software, respectively. Comparing the r values, the correlation between DP-ST-V-Net and ground truth and professional software is better. There is a strong correlation between DP-ST-V-Net and ground truth (r=0.920, P<0.001), and a strong correlation with professional commercial software (r=0.799, P<0.001). As shown in Fig. 11, DP-ST-V-Net achieves a higher correlation between male LVEF and ground truth (r=0.913, P < 0.001) and a higher correlation between female LVEF and ground truth (r=0.914, P < 0.001) than the baseline models. Furthermore, Similar studies were performed on EDV and ESV using the same method, as shown in Fig. 12 and Fig. 13. The EDV and ESV results were calculated between DP-ST-V-Net and ground truth. DP-ST-V-Net and professional commercial software show a better correlation than the other baseline models.

**Table 4**

Correlation r values of clinical parameters on different myocardial ischemia patient data.

| Correlation r Value | Severity | LVEF | ESV | EDV |
|---|---|---|---|---|
| DP-ST-V-Net – | Normal or | 0.9243 | 0.9604 | 0.9543 |
| Ground Truth | Mild | | | |
| | Moderate | 0.9221 | 0.9577 | 0.9512 |
| | Severe | 0.9178 | 0.9601 | 0.9514 |
| DP-ST-V-Net – Commercial Software | Normal or Mild | 0.8144 | 0.8654 | 0.9347 |
| | Moderate | 0.8015 | 0.86 | 0.9316 |
| | Severe | 0.78433 | 0.8632 | 0.9324 |

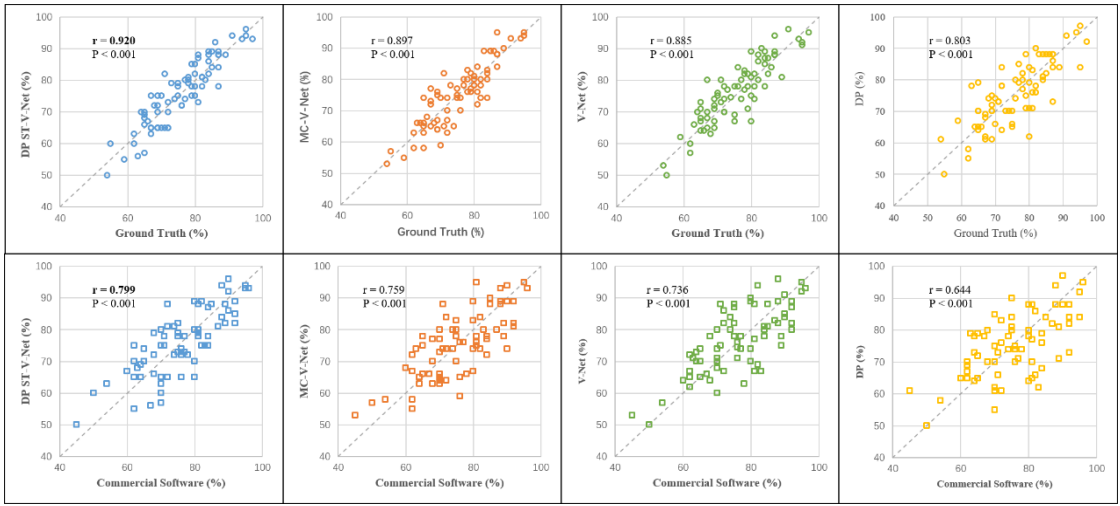

**Fig. 10.** Correlation analysis of calculated LVEF between DP-ST-V-Net, MC-V-Net, V-Net, DP and ground truth and commercial software, respectively.

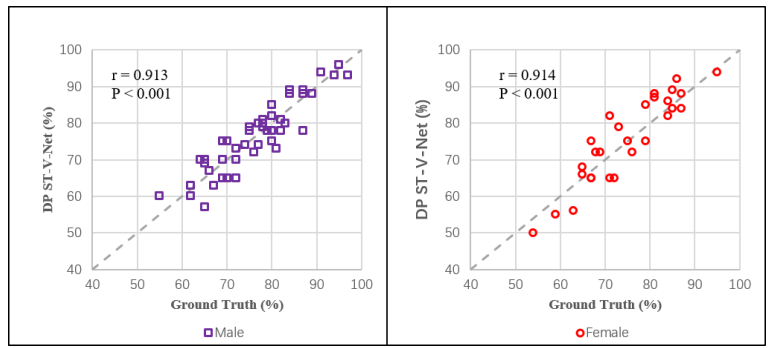

**Fig. 11.** Correlation analysis of DP-ST-V-Net and ground truth to calculate LVEF in 27 males and 18 females.

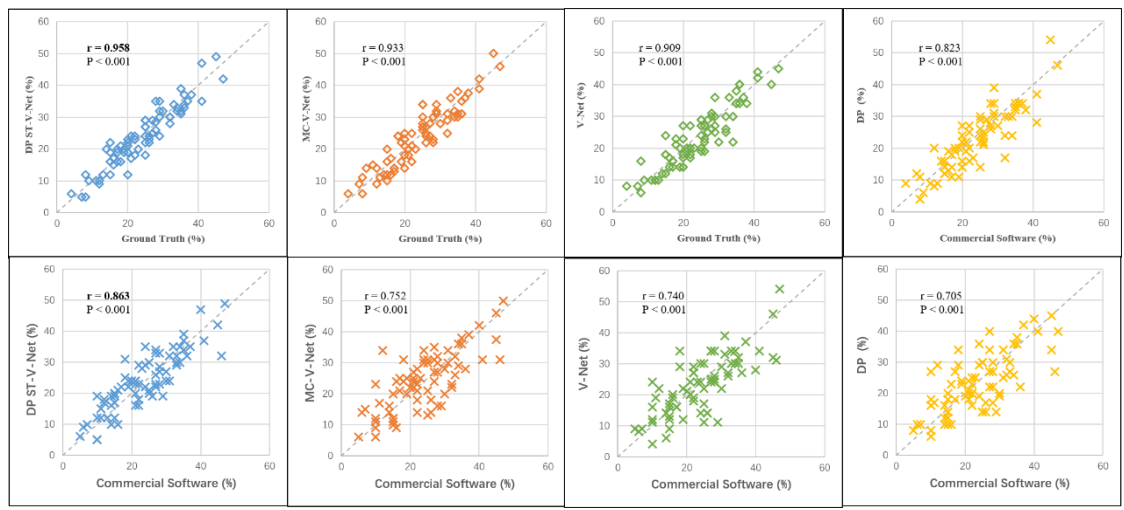

**Fig. 12.** Correlation analysis of the calculated ESV between DP-ST-V-Net, MC-V-Net, V-Net, DP and ground truth and commercial software, respectively.

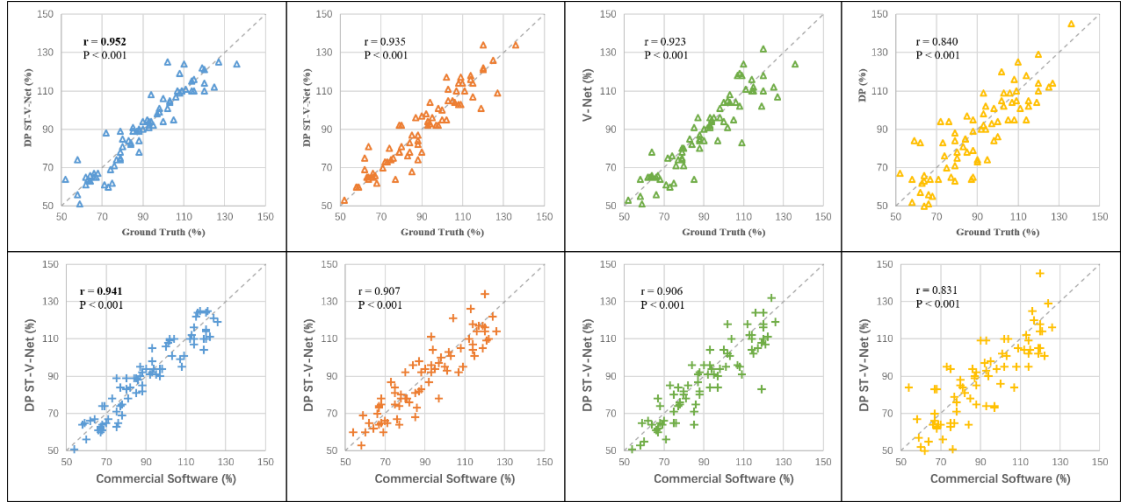

**Fig. 13.** Correlation analysis of the calculated EDV between DP-ST-V-Net, MC-V-Net, V-Net, DP and ground truth and commercial software, respectively.

In Fig. 14, for the 75 stress MPIs, the Pearson correlation coefficients are 0.972 in scar burden; for the 75 rest MPIs, the Pearson correlation coefficients are 0.958 in scar burden, both data are statistically significant.

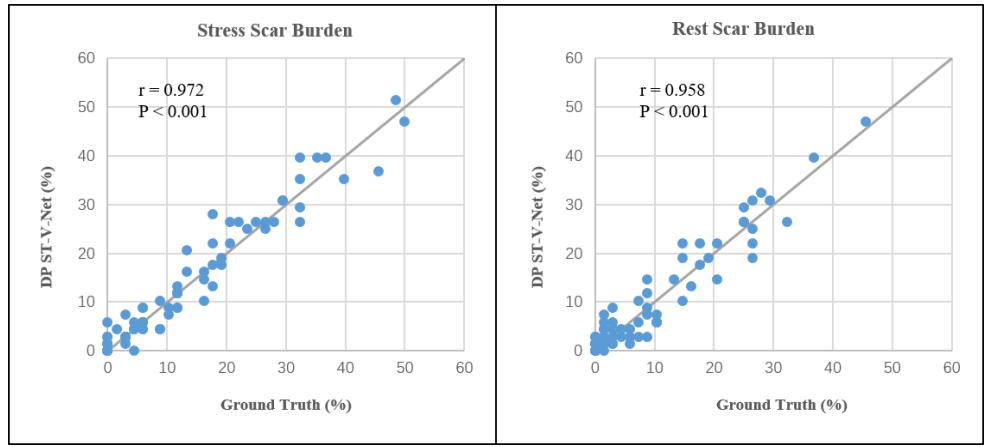

**Fig. 14.** Correlation analysis of scar burden calculated by DP-ST-V-Net and ground truth. A scar sample is defined as less than 50% of the maximum uptake and scar burden is the percentage of the scar samples in the LV myocardium.

## 4. Discussion

### 4.1 Performance analysis

The contours of the proposed method for LV segmentation in cardiac MPS images are in good agreement with the ground truth. As listed in Table 1, the mean DSCs of myocardial contours delineated by DP-ST-V-Net are all greater than 0.95, and the HD values are less than 8 mm (image voxel size: 6.4 mm × 6.4 mm × 6.4 mm). Compared with other MPS segmentation methods in the table, the DSC value of this method is significantly improved, and the HD value is significantly decreased. Ablation experiments are performed to investigate whether the introduction of shape priors and shape deformation modules is effective. As listed in Table 2, by adding two modules in sequence, the performance of the model is gradually improved. Therefore, adding shape prior and shape deformation modules can effectively improve the performance of the model. The results of contour segmentation are similar for the three categories of myocardial ischemic severity, as listed in Table 3, which demonstrates that the DP-ST-V-Net has a high degree of similarity between its output contours and ground truth during segmentation of the myocardium regardless of severity of myocardial ischemia in the patient.

The endocardium tends to be smaller than the LV myocardium and epicardium. Thus, the background region is favored during model output, and the segmented region is ignored. DP-ST-V-Net shows good performance even in recognizing endocardial contours, outperforming most current LV myocardial segmentation methods in gated MPS images. As shown in Fig. 5 and Fig. 6, the DP-ST-V-Net segmentation contours agree well with the ground truth.

**4.2 Clinical analysis**

Quantification of LV function parameters is critical for determining appropriate treatment plans and predicting the occurrence of adverse events [32]. However, when commercialized software commonly used clinically processes MPS images, it generally requires a specialized physician to perform correction or intervention. This is time-consuming and dependent on the subjective experience of the physician. Therefore, a reliable segmentation method is essential for the quantitative analysis of gated MPS images.

Two of the difficulties of LV segmentation in MPS images are avoiding errors caused by extracardiac activity and reduced developer uptake: (1) In cardiac perfusion studies, the reduction of myocardial blood flow and myocardial activity will reduce the uptake of imaging agents, and some myocardial regions cannot be displayed, which affects the identification of myocardial contours. (2) Problems such as right ventricular myocardial hypertrophy will interfere with the heart, resulting in distortion of the myocardial boundary. Applying DP-ST-V-Net performs well on data with different myocardial ischemia severities, as shown in Table 3. The results show that the proposed segmentation algorithm is equally effective regardless of the severity of myocardial ischemia in the patient.

The measurements extracted from this study are clinically significant. Decreased LVEF is associated with heart failure. The increase of LV myocardial volume reflects the damage of myocardium, and diseases such as hypertension, coronary heart disease, diabetes and chronic renal failure can cause the increase of myocardial volume [33]. Changes in the endocardial surface between EDV and ESV can further estimate wall motion. Related studies have demonstrated that existing methods of calculating LV myocardial volume often overestimate small heart volumes and underestimate large heart volumes, and that LVEF is more likely to be overestimated in female patients. In the experimental results, the mean correlation coefficient between measured volume size and volume error in 75 patients is 0.187 ($P=0.262$), which is not statistically significant. In Fig. 10, the correlation coefficient of DP-ST-V-Net for calculating LVEF for males is 0.931 ($P<0.001$) and for females is 0.926 ($P<0.001$), and there is only a minor difference between the correlations. It can be inferred that the DP-ST-V-Net is able to accurately measure LV volume and LVEF regardless of myocardial size and patient's gender.

Validation of the LVEF does not imply validation of the EDV and ESV measurements on which the LVEF is based. For example, when calculating volume ratios for LVEF calculation purposes, the errors determined by EDV and ESV are offset (errors are expected to appear in the same general direction) [8]. Despite these issues, the findings show that the measurements of absolute LV cavity from DP-ST-V-Net are in good agreement with the ground truth results.

For LVEF, the correlation coefficient between the DP-ST-V-Net segmentation results and the commercial program calculation is 0.799. The correlation with EDV is strong ($r=0.941$), and the result with ESV is different ($r=0.863$). From this, it can be inferred that the measurement differences in LVEF are mainly due to ESV. This may be due to different methods of calculating ESV in commercial software. Studies have shown that the correlation between EFs determined using different two commercial software may be around 0.800 [34]. Therefore, the results based on the commercial software should be regarded as only a benchmark rather than a gold standard in the current study.

In addition to measurement of LV volumes and LVEF, scar burden, a common clinical parameter for assessment of LV function [35], is also measured for validation of LV myocardial segmentation. Figure 14 shows that scar burden measured from DP-ST-V-net results agrees with that from the ground truth, suggesting a strong promise of clinical use of our proposed method.

### 4.3 Future study

The V-Net output needs to be refined based on a shape prior. Thus, a reliable shape prior is vital to the model's overall performance. Studies have shown that the DP algorithm has good accuracy in segmenting the LV myocardium. Therefore, it is appropriate to choose the myocardial contours generated by DP as a shape prior. However, it is evident from Table 1 that the DP is slightly less accurate when segmenting the endocardial region. In future studies, other methods are chosen to generate shape priors to compensate for the inaccuracy of DP in identifying endocardium.

In addition, more data from patients with different pathological abnormalities will be included for a more comprehensive evaluation. Data outlined by other physicians (e.g., multiple physician consensus profiles) will be used as training and testing data sets to further evaluate the clinical application of the method. More importantly, our approach will be applied to the measurement of important clinical parameters on MPS [35] and image fusion to improve cardiac resynchronization therapy for heart failure [36] and coronary revascularization for CAD [37].

## 5. Conclusion

Our proposed method achieved a high accuracy in extracting LV myocardial contours and assessing LV function. Future studies will further improve our method and investigate the values of LV parameters measured from our method in a clinical environment.

**Declaration of Competing Interest**

The authors report no declarations of interest.

**Authors' contribution**

**Fubao Zhu**: article proofreading, project administration, and funding acquisition. **Jinyu Zhao**: coding, experiment, and original draft. **Chen Zhao**: methodology, article proofreading. **Shaojie Tang**: methodology, article proofreading. **Jiaofen Nan**: article proofreading. **Yanting Li**: article proofreading. **Zhongqiang Zhao**: data acquisition, clinical evaluation. **Jianzhou Shi**: data acquisition, clinical evaluation. **Zenghong Chen**: data acquisition, clinical evaluation. **Zhixin Jiang**: clinical evaluation, article proofreading and approval. **Weihua Zhou**: project design, project administration, article proofreading and approval.

**Acknowledgment**

This research was funded by the Henan Province Science and Technology Development Plan Project in 2022 [Project Number: 222102210219], National Natural Science Foundation of China [Project Number: 62106233] and Shaanxi Provincial Natural Science Foundation of China [Project Number: 2020SF377].

**References**

[1] Virani, S. S., Alonso, A., Benjamin, E. J., Bittencourt, M. S., Callaway, C. W., Carson, A. P., et al. (2020). Heart Disease and Stroke Statistics-2020 Update: A Report From the American Heart Association. Circulation, 141(9), e139-e596. https://doi.org/10.1161/cir.0000000000000757.

[2] Klocke, F. J., Baird, M. G., Lorell, B. H., Bateman, T. M., Messer, J. V., Berman, D. S., et al. (2003). ACC/AHA/ASNC guidelines for the clinical use of cardiac radionuclide imaging--executive summary: a report of the American College of Cardiology/American Heart Association Task Force on Practice Guidelines (ACC/AHA/ASNC Committee to Revise the 1995 Guidelines for the Clinical Use of Cardiac Radionuclide Imaging). J Am Coll Cardiol, 42(7), 1318-1333. https://doi.org/10.1016/j.jacc.2003.08.011.

[3] Slomka, P., Xu, Y., Berman, D., and Germano, G. (2012). Quantitative analysis of perfusion studies: strengths and pitfalls. J Nucl Cardiol, 19(2), 338-346. https://doi.org/10.1007/s12350-011-9509-2.

[4] Sharir, T., Germano, G., Kang, X., Lewin, H. C., Miranda, R., Cohen, I., et al. (2001). Prediction of myocardial infarction versus cardiac death by gated myocardial perfusion SPECT: risk stratification by the amount of stress-induced ischemia and the poststress ejection fraction. J Nucl Med, 42(6), 831-837.

[5] Friehling, M., Chen, J., Saba, S., Bazaz, R., Schwartzman, D., Adelstein, E. C., et al. (2011). A prospective pilot study to evaluate the relationship between acute change in left ventricular synchrony after cardiac resynchronization therapy and patient outcome using a single-injection gated SPECT protocol. Circ Cardiovasc Imaging, 4(5), 532-539. https://doi.org/10.1161/circimaging.111.965459.

[6] Xu, Y., Kavanagh, P., Fish, M., Gerlach, J., Ramesh, A., Lemley, M., et al. (2009). Automated quality control for segmentation of myocardial perfusion SPECT. J Nucl Med, 50(9), 1418-1426. https://doi.org/10.2967/jnumed.108.061333.

[7] Germano, G., Kavanagh, P. B., Waechter, P., Areeda, J., Van Kriekinge, S., Sharir, T., et al. (2000). A new algorithm for the quantitation of myocardial perfusion SPECT. I: technical principles and reproducibility. J Nucl Med, 41(4), 712-719.

[8] Germano, G., Kiat, H., Kavanagh, P. B., Moriel, M., Mazzanti, M., Su, H. T., et al. (1995). Automatic quantification of ejection fraction from gated myocardial perfusion SPECT. J Nucl Med, 36(11), 2138-2147.

[9] Nakajima, K., Okuda, K., Nyström, K., Richter, J., Minarik, D., Wakabayashi, H., et al. (2013). Improved quantification of small hearts for gated myocardial perfusion imaging. Eur J Nucl Med Mol Imaging, 40(8), 1163-1170. https://doi.org/10.1007/s00259-013-2431-x.

[10] Hambye, A. S., Vervaet, A., and Dobbeleir, A. (2004). Variability of left ventricular ejection fraction and volumes with quantitative gated SPECT: influence of algorithm, pixel size and reconstruction parameters in small and normal-sized hearts. Eur J Nucl Med Mol Imaging, 31(12), 1606-1613. https://doi.org/10.1007/s00259-004-1601-2.

[11] Soneson, H., Ubachs, J. F., Ugander, M., Arheden, H., and Heiberg, E. (2009). An improved method for automatic segmentation of the left ventricle in myocardial perfusion SPECT. J Nucl Med, 50(2), 205-213. https://doi.org/10.2967/jnumed.108.057323.

[12] Petitjean, C., Zuluaga, M. A., Bai, W., Dacher, J.-N., Grosgeorge, D., Caudron, J., et al. (2015). Right

ventricle segmentation from cardiac MRI: A collation study. MEDICAL IMAGE ANALYSIS, 19(1), 187-202. https://doi.org/https://doi.org/10.1016/j.media.2014.10.004.

[13] Peng, P., Lekadir, K., Gooya, A., Shao, L., Petersen, S. E., and Frangi, A. F. (2016). A review of heart chamber segmentation for structural and functional analysis using cardiac magnetic resonance imaging. Magma, 29(2), 155-195. https://doi.org/10.1007/s10334-015-0521-4.

[14] Tavakoli, V., and Amini, A. A. (2013). A survey of shaped-based registration and segmentation techniques for cardiac images. Computer Vision and Image Understanding, 117(9), 966-989. https://doi.org/https://doi.org/10.1016/j.cviu.2012.11.017.

[15] Lesage, D., Angelini, E. D., Bloch, I., and Funka-Lea, G. (2009). A review of 3D vessel lumen segmentation techniques: Models, features and extraction schemes. MEDICAL IMAGE ANALYSIS, 13(6), 819-845. https://doi.org/https://doi.org/10.1016/j.media.2009.07.011.

[16] Hu, Z., Tang, J., Wang, Z., Zhang, K., Zhang, L., and Sun, Q. (2018). Deep learning for image-based cancer detection and diagnosis − A survey. Pattern Recognition, 83, 134-149. https://doi.org/https://doi.org/10.1016/j.patcog.2018.05.014.

[17] Wang, T., Lei, Y., Tang, H., He, Z., Castillo, R., Wang, C., et al. (2020). A learning-based automatic segmentation and quantification method on left ventricle in gated myocardial perfusion SPECT imaging: A feasibility study. J Nucl Cardiol, 27(3), 976-987. https://doi.org/10.1007/s12350-019-01594-2.

[18] Wen, H., Wei, Q., Huang, J.-L., Tsai, S.-C., Wang, C.-Y., Chiang, K.-F., et al. (2021). Analysis on SPECT myocardial perfusion imaging with a tool derived from dynamic programming to deep learning. Optik, 240, 166842. https://doi.org/https://doi.org/10.1016/j.ijleo.2021.166842.

[19] Zhao, C., Xu, Y., He, Z., Tang, J., Zhang, Y., Han, J., et al. (2021). Lung segmentation and automatic detection of COVID-19 using radiomic features from chest CT images. Pattern Recognit, 119, 108071. https://doi.org/10.1016/j.patcog.2021.108071.

[20] Zhao, C., Keyak, J. H., Tang, J., Kaneko, T. S., Khosla, S., Amin, S., et al. (2021). ST-V-Net: incorporating shape prior into convolutional neural networks for proximal femur segmentation. Complex & Intelligent Systems. https://doi.org/10.1007/s40747-021-00427-5.

[21] Zhao, C., Vij, A., Malhotra, S., Tang, J., Tang, H., Pienta, D., et al. (2021). Automatic extraction and stenosis evaluation of coronary arteries in invasive coronary angiograms. Computers In Biology And Medicine, 136, 104667. https://doi.org/10.1016/j.compbiomed.2021.104667.

[22] Ravishankar, H., Venkataramani, R., Thiruvenkadam, S., Sudhakar, P., and Vaidya, V. (2017, 2017//). Learning and Incorporating Shape Models for Semantic Segmentation. Paper presented at the Medical Image Computing and Computer Assisted Intervention − MICCAI 2017, Cham.

[23] Lee, M. C. H., Petersen, K., Pawlowski, N., Glocker, B., and Schaap, M. (2019). TeTrIS: Template Transformer Networks for Image Segmentation With Shape Priors. IEEE Trans Med Imaging, 38(11), 2596-2606. https://doi.org/10.1109/tmi.2019.2905990.

[24] Tang, S., Hung, G., Tsai, S., Wang, C., Li, D., and Zhou, W. (2017). Dynamic programming-based automatic myocardial quantification from the gated SPECT myocardial perfusion imaging, in: The International Meeting on Fully Three-Dimensional Image Reconstruction in Radiology and Nuclear Medicine, Xi'an, 5 China, , pp. 462–467.

[25] Srivastava, N., Hinton, G., Krizhevsky, A., Sutskever, I., and Salakhutdinov, R. (2014). Dropout: A Simple Way to Prevent Neural Networks from Overfitting. Journal of Machine Learning Research, 15, 1929-1958. http://portal.acm.org/citation.cfm?id=2670313.

[26] Wang, C., Tang, S., Tang, H., Rui, Gao., Li, D., and Zhou, W. (2017). A New Method to Automatically Identify Left-ventricular Contours from The Gated SPECT Myocardial Perfusion Imaging, The 14th International Meeting on Fully Three-Dimensional Image Reconstruction in Radiology and Nuclear Medicine.

[27] Jaderberg, M., Simonyan, K., Zisserman, A., and Kavukcuoglu, K. (2015). Spatial transformer networks. Paper presented at the Proceedings of the 28th International Conference on Neural Information Processing Systems - Volume 2, Montreal, Canada.

[28] Bernard, O., Bosch, J. G., Heyde, B., Alessandrini, M., Barbosa, D., Camarasu-Pop, S., et al. (2016). Standardized Evaluation System for Left Ventricular Segmentation Algorithms in 3D Echocardiography. IEEE Trans Med Imaging, 35(4), 967-977. https://doi.org/10.1109/tmi.2015.2503890.

[29] Lum, D. P., and Coel, M. N. (2003). Comparison of automatic quantification software for the measurement of ventricular volume and ejection fraction in gated myocardial perfusion SPECT. Nucl Med Commun, 24(3), 259-266. https://doi.org/10.1097/00006231-200303000-00005.

[30] Ringenberg, J., Deo, M., Devabhaktuni, V., Berenfeld, O., Boyers, P., and Gold, J. (2014). Fast, accurate, and fully automatic segmentation of the right ventricle in short-axis cardiac MRI. Comput Med Imaging Graph, 38(3), 190-201. https://doi.org/10.1016/j.compmedimag.2013.12.011.

[31] Germano, G., Kiat, H., Kavanagh, P. B., Moriel, M., Mazzanti, M., Su, H. T., et al. (1995). Automatic quantification of ejection fraction from gated myocardial perfusion SPECT. J Nucl Med, 36(11), 2138-2147.

[32] Levy, D., Garrison, R. J., Savage, D. D., Kannel, W. B., and Castelli, W. P. (1989). Left ventricular mass and incidence of coronary heart disease in an elderly cohort. The Framingham Heart Study. Ann Intern Med, 110(2), 101-107. https://doi.org/10.7326/0003-4819-110-2-101.

[33] Germano, G., Kavanagh, P. B., Slomka, P. J., Van Kriekinge, S. D., Pollard, G., and Berman, D. S. (2007). Quantitation in gated perfusion SPECT imaging: the Cedars-Sinai approach. J Nucl Cardiol, 14(4), 433-454. https://doi.org/10.1016/j.nuclcard.2007.06.008.

[34] Nakajima, K., Higuchi, T., Taki, J., Kawano, M., and Tonami, N. (2001). Accuracy of ventricular volume and ejection fraction measured by gated myocardial SPECT: comparison of 4 software programs. J Nucl Med, 42(10), 1571-1578.

[35] Zhou, W., and Garcia, E. V. (2016). Nuclear Image-Guided Approaches for Cardiac Resynchronization Therapy (CRT). Curr Cardiol Rep, 18(1), 7. https://doi.org/10.1007/s11886-015-0687-4.

[36] Zhou, W., Hou, X., Piccinelli, M., Tang, X., Tang, L., Cao, K., et al. (2014). 3D fusion of LV venous anatomy on fluoroscopy venograms with epicardial surface on SPECT myocardial perfusion images for guiding CRT LV lead placement. JACC Cardiovasc Imaging, 7(12), 1239-1248. https://doi.org/10.1016/j.jcmg.2014.09.002.

[37] Xu, Z., Tang, H., Malhotra, S., Dong, M., Zhao, C., Ye, Z., et al. (2022). Three-dimensional Fusion of Myocardial Perfusion SPECT and Invasive Coronary Angiography Guides Coronary Revascularization. J Nucl Cardiol. https://doi.org/10.1007/s12350-022-02907-8.